\documentclass[aps,reprint,twocolumn,showpacs,amsmath,amssymb]{revtex4}
\usepackage{epsfig}
\usepackage{dcolumn}
\usepackage{bm}

\begin{document}

\title{Statistics of heat generated in a solvable dissipative Landau-Zener model.}
\author{V.V. Ponomarenko}
\affiliation{Center of Physics, University of Minho, Campus
Gualtar, 4710-057 Braga, Portugal}

\date{\today}

\begin{abstract}
We consider an adiabatic Landau-Zener model of  two-level system
diagonally coupled to an Ohmic bosonic bath of large spectral
width and derive through fermionization its exact solution at a
special value of the coupling constant. From this solution we obtain
the characteristic function of the distribution of energy transferred
to the bath during the evolution of the system ground state
as a functional determinant of a single
particle operator. At zero temperature  this distribution is
further found to be exponential and at finite temperature the
first three moments of the distribution are calculated.

\end{abstract}

\pacs{74.78.Na, 03.65.Yz, 74.50.+r, 05.40.-a}

\maketitle

Recently the interest to dissipative Landau-Zener (LZ) models has
been revived in the context of studying fluctuation relations
\cite{FR1,FR2,FR3,FR4} for work and dissipation in small systems
driven out of equilibrium by external force. The main focus of the
renewed interest is, however, different from the earlier studied
\cite{Ao,disLZ2} effect of a dissipative environment on the LZ
probability of the non-adiabatic transition. Now, it is the
quantum statistics of the energy transferred to the environment.
This interest has grown in connection with the definition and
measurement of the work performed on a small quantum system in the
non-equilibrium process in the quantum version
\cite{QFR1,QFR2,QFR3} of the fluctuation relations.  It has been
suggested \cite{Cr} based on the principle of conservation of
energy to define this work through the heat generated in the
environment. To study this matter a simple, but experimentally
feasible system of superconducting Cooper-pair box driven by a
gate voltage has been considered \cite{PA,SAP} in the regime when
its theoretical description reduces to a dissipative LZ model of
two levels undergoing avoiding crossing and coupled to an Ohmic
bosonic bath. In spite of its simplicity a non-perturbative
quantum solution to this interacting model still remains a
challenging task, in particular, since at low energy the model is
equivalent \cite{L} to an anisotropic Kondo model driven out of
the thermal equilibrium by the time dependent magnetic field or,
equally, to an interacting resonant level model (IRLM) with the
time dependent level energy \cite{me}. Their stationary
equilibrium solution, in general, is available only in the Bethe
ansatz technique \cite{BA}. Its generalization to the stationary
non-equilibrium IRLM of electronic transport is difficult
\cite{OBA1,OBA2} and remains completely unknown to the
non-stationary models.

In this work therefore we consider a special case of this
dissipative LZ model at a particular value of the bath coupling
constant which corresponds to the Toulouse limit of the
anisotropic Kondo model solvable through re-fermionization. In
equilibrium this special case of the Kondo model has been
particular important since it gives a simple but universal
description of the low energy Fermi liquid behavior characterizing
an antiferromagnetic fixed point for the renormalization group
scaling procedure \cite{RG}. Therefore this particular LZ model of
the two-level system will also show the general low energy
properties of the heat distribution generated during the system
evolution. We will construct solution to this model and use it to
calculate the heat distribution in the adiabatic limit when the
system enters and exits the evolution in its ground states. In this limit
the excitations produced in the Ohmic
environment of wide energy spectrum are limited to smaller
energies than the environment spectral width.

\emph{Model} - The LZ model ( also known \cite{LZSM} as the
Landau-Zener-St\"{u}ckelberg-Majorana model) describes transition
of the system between its two states denoted as spin up (down)
$|\uparrow(\downarrow)> $  with the time dependent Hamiltonian
${\cal H}_S(t)= a \, t \,\sigma_z/2+\Delta \sigma_x$. Here
$\sigma_{x(z)}$ are Pauli matrices and the constant sweep velocity
$a(>0)$ regulates crossing of the diabatic energies $\pm a\, t/2$
of the two states coupled by the tunneling amplitude $\Delta$. The
interaction of the system with the environment modeled as a bath
of the harmonic oscillators is introduced by the additional part
of the Hamiltonian $(\hbar=1)$
\begin{equation}
{\cal H}_E=\int \frac{dx}{4 \pi} (\partial_x \phi)^2+U\sigma_z\partial_x \phi(0)
\ ,
\label{he}
\end{equation}
where annihilation and creation operators of the oscillators are
combined into the bosonic chiral field $\phi(x)=\int d \omega
\exp(-i\omega x)\phi(\omega)$. Its Fourier components satisfy
$[\phi(-\omega'),\phi(\omega)]=\delta(\omega-\omega')/\omega $ .
Therefore the spectral function of the bath is defined \cite {Ao}
by the correlator of the coupling operator in Eq. (\ref{he}) as
$J(\omega)=(2U)^2\omega \exp(-\omega/D)$, where the energy cut-off
$D$ is assumed to be much larger than all other energy parameters
in the model, in particular, $D \gg \Delta$. This smooth
exponential cut-off of $J(\omega)$  substitutes for a more
realistic Lorentzian one used in Ref. \cite{PA}.

By applying the unitary transformation ${\cal
U}=\exp(i\phi(0)\sigma_z/2)$ to the sum of both parts of the
Hamiltonian ${\cal U}^+[{\cal H}_S+{\cal H}_E]{\cal U} $ and
making use of the fermionic representation of Pauli matrices
$\sigma_z=2d^+d-1, \sigma_x=\sigma_+ + \sigma_-=d^+\eta+\eta d$,
where $\eta$ denotes an auxiliary  Majorana fermion and $d$ is the
annihilation operator of another fermion we come to the fermionic
description of the model by the time-dependent IRLM Hamiltonian
\begin{eqnarray}
{\cal H}_{F}(t)={\cal
H}_0+at(d^+d-\frac{1}{2})+w(d^+\psi(0)+h.c.)\nonumber
\\
+ \pi(2U-1)\psi^+(0)\psi(0)(2d^+d-1) \ , \label{hf}\\
{\cal H}_0=-i\!\! \int\! dx \psi^+(x)
\partial_x \psi(x) \ , \nonumber
\end{eqnarray}
where the chiral Fermi field $\psi(x)$ stands for $
\psi(x)=\sqrt{\frac{D}{2\pi} }\eta e^{i\phi(x)}$, the Fermi sea of
occupied fermion states is defined by the zero chemical potential,
and the tunneling amplitude $w$ is $w=\Delta \sqrt{2 \pi/D}$. The
density of the Fermi states also undergoes the same exponential
cut-off at large absolute values of their energies as the bosonic
bath modes. In this formulation of the problem the case of the
system being at $t=-\infty$ in the ground state, which we will
study below, corresponds to the filled resonant level entering the
Fermi sea from its underneath. The exponential boundaries $\pm D$
of the Fermi states energies do not appear in the Hamiltonian
(\ref{hf}) directly and in this form it describes only evolution
of the states with energies deep inside the energy band. In the
stationary case $a=0$ this description is accurate  since the
tunneling rate $\gamma=w^2/2$ is small $\gamma \ll D$. Although in
our time dependent IRLM the resonant level traverses the whole
fermion energy band the use of this Hamiltonian is still justified
if the tunneling in and out of the resonant level vanishes quickly
enough far from the Fermi level and the energies of the Fermi sea
excitations remain much less than $D$.

\emph{Fermion model solution} - We will find solution to the model
described by Eq. (\ref{hf}) at $U=1/2$, when the contact
interaction between the resonant level and the Fermi sea vanishes.
Then the equations of motion for the fermion operators can be
written as follows:
\begin{eqnarray}
i \partial d(t) =at d(t)+\frac{w}{2}(\psi_{in}(t)+\psi_{out}(t)) \ ,
\label{df}\\
\psi_{out}(t)=\psi_{in}(t)-i w d(t) \ , \nonumber
\end{eqnarray}
where the incoming and outgoing fermions $\psi_{in(out)}$ describe
the chiral propagation of $\psi$ on both sides from the resonant
level as
$\psi(x,t)=\theta(-x)\psi_{in}(t-x)+\theta(x)\psi_{out}(t-x)$.
Further solving the linear differential equation we express the
outgoing fermion operators at time $t$ through the incoming
operators at earlier times starting from the initial one $t_0$
\begin{eqnarray}
\psi_{out}(t)=\psi_{in}(t)-2\gamma e^{-iE(t)} \int^t_{t_0}
d\tau e^{\gamma(\tau-t)+iE(\tau)} \psi_{in}(\tau) \nonumber \\
-i w e^{\gamma(t_0-t)} e^{i(E(t_0)-E(t))} d(t_0) \ , \ \ E(t)=at^2/2    \ \
\label{outin}
\end{eqnarray}
Here
$t_0$ can be chosen as early as the entrance time of the level
into the fermion band: $t_0 \approx -D/a$. Then under assumption
that the traversal time $D/a$ is much larger than the tunneling
time and $\gamma D/a \gg 1$,  decay
of the resonant level state makes contribution from the last term in
Eq. (\ref{outin}) negligibly small
at finite time and the low limit of integration in the second term
may be drawn to $-\infty$. In the resultant relation between the
incoming and outgoing fermions it is convenient to represent both
Fermi fields as $\psi_\alpha(t)=\!\int \! dk
\exp(-i[kt-k^2/(2a)])c_\alpha(k)/\sqrt{2\pi} \, , \alpha=in,out$.
In this representation the $S$ matrix relating incoming and
outgoing plane waves in this scattering problem $c_{out}(k)=\int
dk' S(k,k') c_{in}(k') $ follows from Eq. (\ref{outin}) in the
simple form:
\begin{equation}
S(k,k')=\delta(k-k')-\frac{2\gamma}{a}\theta(k-k')e^{\gamma
(k'-k)/a} \ . \label{s}
\end{equation}
This $S$ matrix is unitary in the absence of the energy band
restrictions on the fermion energies $k$. Therefore it properly
describes the low energy scattering under the same assumption of
the large enough traversal time when we can neglect the
exponentially small probability $\exp(-2 \gamma D/a)$ for a
low energy fermion to reach the energy band boundary.
Since $\gamma D/a=\pi \Delta^2/a$
this is also the LZ probability of the
non-adiabatic transition, which at zero temperature is not
affected \cite{Ao,disLZ2,Shytov} by the diagonal coupling of the
system to the bath in Eq. (\ref{he}).

\emph{Characteristic function of heat} - Under this adiabatic
assumption we can limit our consideration only to the evolution of
the Fermi sea. Then the characteristic function of the
distribution of its energy excitations $\chi(\lambda)$ can be
expressed through the $S$ matrix following the lines of derivation
\cite{Klitch1} of the Levitov-Lesovik formula \cite{LL} for the
full counting statistics of charge transfer. The result comes as a
determinant of the operator acting in the one-particle Hilbert
space:
\begin{equation}
\chi(\lambda)=det\{1+n_F(e^{-i\lambda h_0}S^+ e^{i\lambda
h_0}S-1)\} \ , \label{chi1}
\end{equation}
where $h_0$ stands for the one-particle Hamiltonian operator
$h_0(k,k')=k \delta(k-k')$ and the operator $n_F$ is defined by
the correspondent Fermi-Dirac distribution function.
As follows from its derivation
and properties the S matrix (\ref{s}) implies the constant
density of the fermion states and hence the infinite depth of the
Fermi sea.
However, the functional determinant is well defined only for the
operator whose difference from the identity is a trace class
operator. Although we give below its general proper regularization
the direct use of the expression Eq.(\ref{chi1}) is also
convenient and possible, if we impose restriction on occupation of
the fermion states below some energy $-W$ through introduction of
an artificial filling factor $\rho(k)=\exp(-|k|/W)$, which will be
lifted at the end of the calculations as $W \rightarrow -\infty$.
In this way all moments of the heat distribution can be found by
taking derivatives of the function
\begin{equation}
\ln\chi(\lambda)=tr\{ \ln[1+n (e^{-i\lambda h_0}S^+ e^{i\lambda
h_0}S-1)]\} \ \label{lnchi1}
\end{equation}
with respect to $i\lambda$ at $\lambda=0$, where $n=\rho n_F$ is
the initial one-particle density operator diagonal in the energy
representation and $tr$ assumes the uniform summation over all $k$
energy states. Then the first derivative gives the average of the
heat generated in the environment as
\begin{equation}
<Q>=tr\{n (S^+ h_0 S-h_0)]\}=tr\{(S n S^+ -n)h_0]\} \ . \label{Q1}
\end{equation}
Both expressions for the average heat in (\ref{Q1}) are
equivalent for the trace convergent density operators, though the
second one permits lifting the filling factor restriction because
it distinguishes contributions from the low and high energy
excitations of the Fermi sea. Indeed, the variation of the
one-particle density operator $\Delta n \equiv S n S^+-n$ is equal
to
\begin{eqnarray}
\Delta n(k,k')=\frac{2 \gamma}{a}e^{-\gamma |k-k'|/a} \Delta n_\|(\min\{k,k'\}) \ ,
\label{dn}\\
\Delta n_\|(p)=-n(p)+\frac{2 \gamma}{a}\int^p_{-\infty} d p' n(p')e^{2\gamma (p'-p)/a}  \ . \nonumber
\end{eqnarray}
At zero temperature the substitution of $n=\rho n_F$ in Eq. (\ref{dn}) leads to
the following expression
\begin{equation}
\Delta n=-\frac{2 \gamma \theta(-k_<)}{2 \gamma W+a}
e^{-\frac{\gamma}{a} |k-k'| +\frac{k_<}{W}} + \frac{4 \gamma^2}{a}
\frac{W \theta(k_<)}{2 \gamma W+a}e^{-\frac{\gamma}{a}(k+k')} \ ,
\label{dnb}
\end{equation}
where $k_<=\min\{k,k'\} $. It shows that any large energy cut-off
$W$ of the filling factor still produces some variations of
the density  deep inside the Fermi sea, which compensate its
variations at small positive energies to insure the particle conservation:
$tr \Delta n=0$, Indeed,
the level rising into the Fermi sea with the unfilled states
below it is empty. On the other hand, by drawing first the cut-off
$W$ to the infinity in Eq. (\ref{dnb}) we eliminate all density
variations at negative energies and the density variation operator
becomes
\begin{equation}
\Delta n_F(k,k') = \frac{2 \gamma}{a}
\theta(k)\theta(k')e^{-\gamma (k+k')/a}
\ , \label{dng}
\end{equation}
but with $tr \Delta n_F=1$. This density variation operator
describes evolution of the Fermi sea caused by the filled level
rising into it and bringing an additional fermion when the system
is in the ground state at $t=-\infty$. In this case we find from
Eq. (\ref{dng}) that under the adiabatical assumption all
excitations produced in the Fermi sea have energies much less than
$D$, which is consistent with our use of the S-matrix (\ref{s}).

At finite temperature and $n=n_F$ in Eq. (\ref{dn}) the function
$\Delta n_\|$ defining the diagonal matrix elements of the density
variation operator can be found through Laplace transformation as
\begin{equation}
\Delta n_{\|}(k) = \frac{1}{2 \pi i} \int_C \frac{ds e^{-sk}  \pi T s}
{\sin(\pi T s)(\frac{2 \gamma}{a}-s)}
\ , \label{dndiag}
\end{equation}
where the contour $C$ coincides with the imaginary axis infinitely
shifted to the right. We confirm from Eqs. (\ref{dn},\ref{dndiag})
that $tr \Delta n_F=1$ does not depend on temperature and so does
the average heat $<Q>=a/(2\gamma)=Da/(2\pi \Delta^2)$ in Eq.
(\ref{Q1}).

Generalizing this method we would obtain the whole distribution of
the heat produced during the evolution of the system ground state
if we managed to transform
the general expression (\ref{chi1}) for the characteristic
function into the form, which permits lifting the filling factor restriction.
This can be done by using a regularization procedure
similar to that developed  in Refs. \cite{Klitch2} and \cite{MA}
for calculation of the charge transfer statistics in transport
problems. To implement it we multiply the determinant in Eq. (\ref{chi1})
from the left and from the right by the mutually canceling factors
$det\exp\{i\lambda n_F h_0\}$ and $det\exp\{-i\lambda (n_F
h_0)_S\}$, respectively, where we denote $S^+(n_F h_0)S \equiv (n_F h_0)_S
$. The result can be brought into the form:
\begin{eqnarray}
\chi(\lambda)=det\{e^{i\lambda n_F h_0}(1-n_F)e^{-i\lambda (n_F
h_0)_S}  \label{chi2}\\
+e^{-i\lambda (1-n_F) h_0}n_F(e^{i\lambda ((1-n_F) h_0)_S}\} \ ,
\nonumber
\end{eqnarray}
which remains well defined for the infinitely deep Fermi sea without
any additional restrictions.
We further demonstrate this with the zero temperature calculations.

\emph{Zero temperature heat distribution} - Since at zero
temperature the density operator $n_F$ becomes a projector
operator, the characteristic function in Eq. (\ref{chi2})
transforms into the following one:
\begin{eqnarray}
\chi(\lambda)=det\{1+S(1-n_F)S^+n_F(e^{-i\lambda h_0}-1) \nonumber
\\
+Sn_F S^+ (1-n_F)(e^{i\lambda h_0}-1)\} \ . \label{chi3}
\end{eqnarray}
Further substitution here $Sn_FS^+=n_F+\Delta n_F$ with $\Delta n_F$
from Eq. (\ref{dng}) makes it possible to calculate logarithm of
the determinant in Eq. (\ref{chi3}) as follows:
\begin{equation}
\ln \chi(\lambda)=\ln(1+\frac{ia\lambda}{2\gamma-ia\lambda}) \ .
\label{lnchig}
\end{equation}
Its $l$'th derivative with respect to  $i\lambda$ at $\lambda=0$ or the $l$'th
order semi-invariant defines the correspondent reduced correlator
$<<Q^l>>$ of the energy dissipated during the ground state evolution of
the two-level system
as $<<Q^l>>=(l-1)!(a/2\gamma)^l$.

Fourier transformation of the characteristic function in Eq.
(\ref{lnchig}) gives the exponential distribution for the
dissipated energy
\begin{equation}
P(Q)=\theta(Q)\frac{2\gamma}{a}e^{-2\gamma Q/a} \ . \label{pg}
\end{equation}
This distribution coincides with the heat distribution derived
\cite{PA} from solution of the master equation describing the same
model. The derivation is based on calculation of the probability
to find the resonant level positioned at the energy $Q$ to be
occupied. Therefore the coincidence is expected at large
dissipated energies but not at small $Q$,  where the master
equation solution for the probability becomes incorrect. Notice
also that the distribution in Eq. (\ref{pg}) assumes the strictly
adiabatic transition because of its normalization.

\emph{Second and third semi-invariants of the heat distributions}
- We will use Eq. (\ref{lnchi1}) to find the second and third
derivatives of $\ln \chi$ at $\lambda=0$. The second derivative
comes as follows:
\begin{equation}
(-i\partial_\lambda)^2\ln \chi=tr\{n \Delta h (1-n)\Delta h \} \ .
\label{q2}
\end{equation}
Here $\Delta h = S^+ h_0 S-h_0$ is the operator of the
one-particle energy variation. Substituting the $S$ matrix from
Eq. (\ref{s}) we find its kernel to be equal to
\begin{equation}
\Delta h(k,k')=e^{-\gamma |k-k'|/a} \ . \label{dh}
\end{equation}
Making use of it we transform the right side of Eq. (\ref{q2})
into a double integral over the energies, where the filling factor
cut-off $W$ can be drawn to the infinity. After we put $n=n_F$ in
Eq. (\ref{q2}) only the particle states close to the Fermi level
contribute to the integral. The integration gives us the second
reduced correlator of the heat as
\begin{equation}
<<Q^2>>=(\frac{a}{2 \gamma})^2+2T^2\psi'(1+\frac{2 \gamma T}{a}) \
, \label{q2g}
\end{equation}
where $\psi' $ is the first derivative of the di-gamma function.
From the known asymptotics of this function it follows that
$<<Q^2>>=(a/2 \gamma)^2+2 \zeta(2) T^2 $  at small temperature
($\zeta(x)$ is the zeta function) and $<<Q^2>>=(T a/\gamma)
(1+O(a^2/(2 \gamma T)^2))$ at large one.

The result of our calculation of the third derivative can be first
brought into the following form:
\begin{eqnarray}
(-i\partial_\lambda)^3\ln \chi=tr\{n \Delta h (1-n)\Delta h
(1-n)\Delta h \} \label{q3}\\
-tr\{n \Delta h n\Delta h (1-n)\Delta h \}+tr\{n \Delta h [h_0,
\Delta h] \} \ . \nonumber
\end{eqnarray}
Here the first two terms remain finite after substitution of the density matrix $n=n_F$
and cancel each other because of the particle-hole symmetry.
The third term however is ill-defined. To
regularize it consistently with Eq. (\ref{chi2}) we substitute the
identity operator in the form $1=n+(1-n)$ between the first $\Delta h$
operator and the commutator. Since $tr\{n \Delta h n [h_0,
\Delta h] \}=0 $ we
find the third reduced correlator of the heat to be equal to:
\begin{eqnarray}
<<Q^3>>=\!\!\int \!\! d k \, n_F(k)\!\!  \int\!\! d k' (1-n_F(k'))
e^{-2 \frac{\gamma}{a}|k-k'|} \nonumber\\\times(k'-k)
=\frac{1}{4}(\frac{a}{\gamma})^3 \, . \, \  \label{q3g}
\end{eqnarray}
It does not depend on the temperature and coincides with the zero
temperature expression found above.

In conclusion, we have considered the adiabatic LZ model of
evolution of the two-level system diagonally coupled to an Ohmic
bosonic bath of large spectral width $D$ and derived through
fermionization its exact solution at the coupling constant
$U=1/2$. From this solution for the system starting evolution in
the ground state and generating only bosonic excitations of the
energies much less than $D$  we have obtained the characteristic
function of the distribution of heat energy $Q$ transferred to the
bath as a functional determinant of a single particle operator.
The determinant has been used to find that the distribution is
exponential at zero temperature and to calculate its first three
moments at finite  temperature. The Fermi liquid behavior of this
particular model is common for the LZ model at arbitrary $U$,
which means that at low energy the heat distribution is an integer
function of $Q$ with the leading linear decrease and has a $T^2$
temperature growth of its dispersion.

\emph{Acknowledgment} - The author is grateful to Dmitri Averin
and Jukka Pekola for bringing up this problem to his attention and
for useful discussions inspiring this work. The work was supported
by FCT(Portugal) and by the Seventh Framework Programme of the
European Commission under Research Fellowship SFRH/BI/52154/2013
and through the project INFERNOS (Grant Agreement No. 308850).

\end{document}